\documentclass[balance,upint,subscriptcorrection,varvw,mathalfa=cal=boondoxo,spanish,french,vietnamese,russian,greek,pdf-a,colorlinks,nofoot]{asmeconf}

\hypersetup{%
	pdfauthor={Amirtaha Taebi},
	pdftitle={2023 ASME IMECE CTI},                  
	pdfkeywords={ASME conference paper, LaTeX template, BibTeX style},
	pdfsubject = {Describes the asmeconf LaTeX template},			  
	pdflicenseurl={https://ctan.org/pkg/asmeconf},
}

\begin{document}




\title{Effect of Measurement Location on Cardiac Time Intervals Estimated by Seismocardiography} 
 
%
%
%

\SetAuthors{%
	Aysha J.\ Mann\affil{1}, 
	Bahram Kakavand\affil{2}, 
	Peshala Thibbotuwawa Gamage\affil{3}, 
	Amirtah\`a Taebi\affil{1}\CorrespondingAuthor{ataebi@abe.msstate.edu}
	}

\SetAffiliation{1}{Mississippi State University, Mississippi State, MS }
\SetAffiliation{2}{Nemours Children's Hospital, Orlando, FL}
\SetAffiliation{3}{Florida Institute of Technology, Melbourne, FL}


\maketitle



\keywords{Seismocardiography, cardiac time intervals, heart rate, left ventricular ejection time, pre-ejection period, electromechanical systole}


\begin{abstract}

Cardiac time intervals (CTIs) are important parameters for assessing cardiac function and can be measured using non-invasive methods such as electrocardiography (ECG) and seismocardiography (SCG). It is widely accepted that SCG signals, when measured from various locations on the chest surface, exhibit distinct temporal and spectral characteristics. In that regard, the goal of this study was to determine the effect of the SCG measurement location on estimating SCG-based CTIs. For this purpose, ECG, SCG, and phonocardiography (PCG) signals were acquired from fourteen healthy adult subjects, both male and female (Age: 23.5 ± 5.16 years old). Subjects laid still in a supine position and were instructed to breathe normally. Data was recorded for 2 minutes and external noise, if any, was noted and removed. For ECG measurement, Einthoven's triangle was used by placing three ECG electrodes under the left and right clavicle, and the right lower abdomen. For SCG, three tri-axial accelerometers were attached on the top, middle, and bottom of the sternum with double-sided tape. In this study, only the dorsoventral components of the SCG signals were analyzed. Using Pan-Tompkin’s algorithm, ECG R peaks and their temporal indices were found. Then, a custom-built algorithm in MATLAB was developed to estimate heart rate (HR) from ECG (HR\textsubscript{ECG}) and SCG (HR\textsubscript{SCG}) signals. Furthermore, SCG fiducial points and CTIs were defined based  on the literature and estimated from the SCG signals recorded from different sternal locations. These fiducial points and CTIs include the temporal indices of aortic valve opening, aortic valve closure, and R-R interval, as well as pre-ejection period, left ventricular ejection time, and electromechanical systole. The average and correlation coefficient (R\textsuperscript{2}) of the CTIs and HRs derived from all three locations were compared. Mean difference and standard deviation were analyzed for the CTIs and their respective sensor location. Results showed that the average R\textsuperscript{2} values between HR\textsubscript{ECG} and HR\textsubscript{SCG} were 0.9930, 0.9968, and 0.9790 for the top, middle, and bottom sternal locations, respectively. In addition, results demonstrated that SCG-based CTIs varied with the SCG measurement locations. In conclusion, these results highlighted the importance of establishing consistent research and clinical protocols for reporting CTIs based on SCG. This work also calls for further investigation into comparing estimated CTIs with gold-standard methods such as echocardiography and 4D cardiac computed tomography. This will help determine the SCG measurement location that provides the most accurate CTI estimations which in turn can improve the accuracy of SCG-based cardiovascular disease diagnosis algorithms.
\end{abstract}


\begin{nomenclature}[1.5cm]
\EntryHeading{Roman letters}
\entry{$t$}{time instant [s]}
\entry{HR}{Heart rate [beats per minute, bpm]}

\EntryHeading{Superscripts and subscripts}
\entry{i}{i\textsuperscript{th} cardiac cycle}
\entry{R}{R peak of the electrocardiogram}

\EntryHeading{Abbreviations}
\entry{AC}{Aortic valve closure}
\entry{AC}{Aortic valve opening}
\entry{CTI}{Cardiac time interval}
\entry{CVD}{cardiovascular disease}
\entry{ECG}{Electrocardiography}
\entry{LVET}{Left ventricular ejection time}
\entry{PCG}{Phonocardiography}
\entry{PEP}{Pre-ejection period}
\entry{QS2}{Electromechanical systole}
\entry{RRI}{Interval between two consecutive R peaks on electrocardiogram}
\entry{SCG}{Seismocardiography}
\end{nomenclature}


\section{Introduction}
Cardiovascular diseases (CVDs) are not only the leading cause of death in the United States but also in the world \cite{okwuosa2016worldwide}. Early diagnosis and continuous monitoring are critical in managing CVDs and reducing the risk of complications \cite{lin2021reducing, mann2022heart}. Recent advances in noninvasive methods for diagnosing CVDs have shown promise in improving patient outcomes \cite{centracchio2022detection}. One approach to evaluating cardiac health is through the analysis of cardiac time intervals (CTIs), which provide valuable information on cardiac mechanics \cite{zakeri2020repeatability}.  CTIs have the potential to help identify a range of CVDs, including chronic myocardial disease, coronary artery disease, myocardial infarction, arterial hypertension, and hypovolemia \cite{tavakolian2016systolic}. 

Methods that are commonly used in determining CTIs involve imaging methods like magnetic resonance imaging and echocardiography \cite{dehkordi2019comparison}. For example, echocardiography is a medical imaging modality that provides real-time imaging of the heart motion through different points in a cardiac cycle which can lead to a quicker and more accurate detection of cardiac abnormalities \cite{alsharqi2018artificial}. However, these imaging techniques are time-consuming, expensive, and require a professional clinician to acquire images.  Additionally, continuous data acquisition would be a challenge. 

Noninvasive methods such as electrocardiography (ECG) and seismocardiography (SCG) have been also used independently or jointly to estimate the fiducial points of such cardiac events and CTIs \cite{cook2022body}. ECG is the measurement of the electrical activity of the heart and is measured using electrodes placed on the body surface. On the other hand, SCG measures vibrations, in the right-to-left, head-to-foot, and dorsoventral directions (i.e., $x$, $y$, and $z$-axis), along the chest well that are caused by cardiac mechanisms including the opening and closing of heart valves, the filling and contraction of the four heart chambers, and blood flow ejection \cite{mann2022heart, dehkordi2019comparison, ha2020contactless, berkaya2018survey, hu2014physiological, taebi2017time, cook2022body, taebi2019recent, sorensen2018definition, di2013beat}.

The QRS complex is a significant event when analyzing ECG signals. It represents the ventricular depolarization that occurs prior to contraction and is known as the large “spike” in the ECG signal \cite{berkaya2018survey}. Since ECG is a gold standard method in cardiac monitoring, the QRS complex, more specifically the R wave, is commonly used as a reference for other signals when estimating CTIs. With that said, in this specific study, ECG, phonocardiogram (PCG), and SCG signals are used to estimate the cardiac time intervals. In regard to the SCG signals, only the $z$-axis is being analyzed because previous work mainly focused on estimating fiducial points from this SCG component \cite{zakeri2020repeatability, sorensen2018definition}. The following fiducial points on a SCG signal are discussed in the literature: the peak of atrial systole, mitral valve closure, isovolumic movement, aortic valve opening (AO), isotonic contraction, the peak of the rapid systolic ejection, aortic valve closure (AC), mitral valve opening, and the peak of the rapid systolic filling \cite{centracchio2022detection}. Specifically, for this study, the CTIs that are being evaluated are based solely on the timings of AO and AC \cite{dehkordi2019comparison}. The main CTIs that are being considered in this study are R-R interval (RRI), pre-ejection period (PEP), left ventricular ejection time (LVET), and electrotechnical systole (QS2), which is also known as total systolic time.

The PEP is defined as the time interval between the R wave on an ECG waveform and the second peak of the SCG signal in the systolic phase \cite{sorensen2018definition}. PEP reflects the control of left ventricular contractility, or speed of contraction, and is independent of heart rate (HR) \cite{di2013wearable}.  A higher PEP indicates a slower contraction speed \cite{di2013wearable, dehkordi2019comparison}. LVET, the systolic phase of the left ventricle ejecting blood to the aorta, is defined as the time between the AO and AC within a single cardiac cycle. LVET also reflects contractility; however, it is influenced by HR unlike PEP \cite{di2013wearable}. The ratio between PEP and LVET (PEP/LVET) is commonly used as a further reflection of contractility \cite{zakeri2020repeatability, dehkordi2019comparison, ha2020contactless, di2013wearable, buxi2017systolic, alhakak2021significance}. Previous work showed that PEP/LVET ratio trends can indicate certain cardiac dysfunction such as mitral stenosis, heart failure, left ventricular disease, and myocardial infarction \cite{tavakolian2016systolic}. Finally, QS2, the total time of heart contraction, is the time interval between the R wave on an ECG waveform and the AC \cite{dehkordi2019comparison, di2013beat}. Since the aforementioned CTIs all occur during the systolic phase of the cardiac cycle, these parameters are also called systolic time intervals.

Previous studies have shown that SCG analyses are effective in determining mechanical functions and CTIs. One study showed that SCG had 86\% accuracy in determining PEP when compared to the gold-standard reference of an ECG. Using the same gold-standard reference, SCG had a 90\% accuracy when determining QS2 \cite{dehkordi2019comparison}. However, it is well known that SCG signals vary based on location along the chest surface \cite{taebi2019recent, ha2019chest}. Therefore, the aim of this study is to determine whether the location of the SCG sensor has an impact on the accuracy of SCG-based CTI estimations. By analyzing the influence of sensor placement on CTI measurements, this study aims to provide insights into optimizing the use of SCG sensors for non-invasive cardiovascular monitoring.


\section{Materials and Methods}
\subsection{Participants}
The study protocol was approved by the Institutional Review Board of Mississippi State University. Data was collected from fourteen healthy human subjects who gave their full informed consent and reported no history of any cardiovascular diseases. The subjects, which consisted of both male and female sexes, were 23.5 ± 5.16 years old (body mass index: 23.93 ± 4.07 kg/m\textsuperscript{2}).

\subsection{Experimental Setup}
Subjects were instructed to remove any chest hair, and a razor was provided if needed. Additionally, upper-body clothing, undergarments, jewelry, and electronics were removed. Modesty coverings were offered for all subjects, regardless of sex. Subjects were then instructed to lay in a supine position, remain still, and breathe normally. Signals were monitored carefully, and if the quality was desirable, data was acquired for ~2 minutes. External noise such as stomach growling, coughing, or sneezing was noted to be removed from the data before the data being processed.

\begin{figure}
\centering\includegraphics[width=1\linewidth]{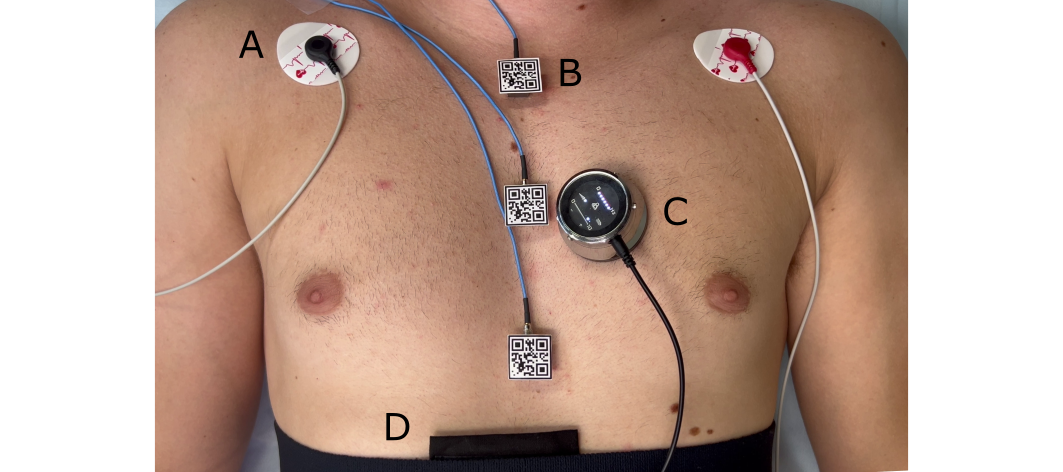}
\caption{Sensor placements on human subjects. The sensors include A) a 3-lead ECG, B) three tri-axial accelerometers, C) an electronic stethoscope, and D) a respiration monitor sensor.}\label{fig:sensors}
\end{figure}

Three tri-axial accelerometers (356A32, PCB Piezotronics, Depew, NY) were placed on the manubrium, left sternal border next to the fourth costal notch, and xiphoid process via double-sided tape (Fig.~\ref{fig:sensors}). The accelerometer outputs were amplified with a gain factor of 100 via a signal conditioner (482C, PCB Piezotronics, Depew, NY). For gold-standard ECG signals, the electrodes were placed under the left and right clavicle, and the right lower abdomen based on Einthoven's triangle configuration. Prior to placing the electrodes, isopropyl alcohol wipes were used to remove any dirt and oil that could influence the electrical activity readings. An electronic stethoscope (Thinklabs One, Thinklabs, Centennial, CO) was used to measure PCG signals and was placed on the fourth left intercostal space, just to the right of the middle accelerometer. Lastly, the respiratory cycle was monitored by a piezo transducer (RM-BIO, iWorx Systems, Inc., Dover, NH) placed around the upper abdomen where the diaphragm is located. The ECG and respiration monitor wires were connected to a recording module (iWire-B3G, iWorx Systems, Inc., Dover, NH). This recording module, electronic stethoscope, and accelerometers were plugged into the data acquisition system (416, iWorx Systems, Inc., Dover, NH) to record the data with a sampling frequency of 5000 Hz. All sensor placements remained consistent on all the subjects, regardless of sex. The placements of all the sensors on one of the participants are presented in Fig.~\ref{fig:sensors}. Once data acquisition began, a tap on the electronic stethoscope marked the beginning of the measurement session. After ~2 minutes passed, another tap indicated the end of the measurement session. The respiration data was not used in the current study.


\subsection{Data Analysis}
\subsubsection*{Preprocessing}
MATLAB (R2022a, The MathWorks, Inc., Natick, MA) was used for signal analysis. In this study, only the dorsoventral components ($z$-axis) of the SCG signals were considered for CTI estimation. Using the extreme maximum in the PCG signals, corresponding to the taps at the beginning and end of the measurement sessions, all signals were trimmed only to have the data that was between the taps. Through this approach, only the relevant data was retained for further analysis, thereby minimizing any potential artifacts that could arise from extraneous noise or interference. Finally, digital filters were designed to remove noise from the ECG and SCG signals. For the ECG signals, a bandpass filter with cutoff frequencies of 0.5 and 40 Hz was applied. A 1-40 Hz bandpass filter was also applied to all SCG signals.

\subsubsection*{Heart Rate Estimation}
The Pan-Tompkin algorithm was used to find the temporal indices of the ECG R peaks. The gold-standard heart rate (HR\textsubscript{ECG}), in beats per minute (bpm), was then estimated from the R peak indices using Eq.~\ref{eqn:hrecg}.

\begin{equation}\label{eqn:hrecg}
HR_{ECG}^i = \frac{1}{t_R^{i+1}-t_R^i} \times 60
\end{equation}
where $t_R^i$ is the time instant of the i\textsuperscript{th} R peak. Similarly, SCG-based HR (HR\textsubscript{SCG}) was estimated by replacing the time instant of the i\textsuperscript{th} R peak with the time instant of the second peak of the SCG1, which is defined as the location of AO. SCG1 refers to the part of an SCG cycle that corresponds to the first heart sound \cite{taebi2019recent}.

\subsubsection*{Fiducial Points and Cardiac Time Intervals}
\begin{figure}[h!]
\centering\includegraphics[width=1\linewidth]{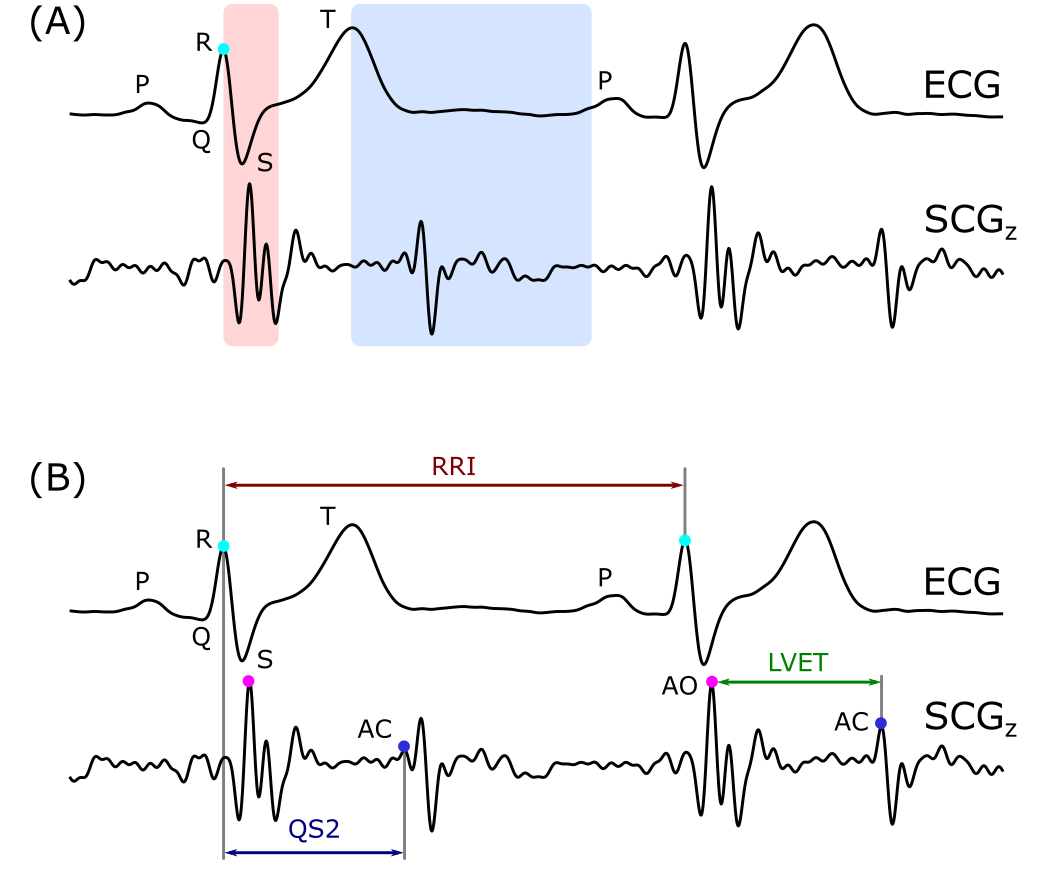}
\caption{(A) ECG-based Search windows to find SCG signal fiducial points. (B) Target fiducial points and CTI\NoCaseChange{s} in this study.}\label{fig:searchWindows}
\end{figure}

Given the known variability of SCG signals based on sensor placement, the signals recorded from the three different sensor locations in this study exhibit noticeable morphological differences. As such, the study utilized specific search windows on the SCG waveforms to obtain accurate CTI measurements, rather than relying on the identification of specific peak heights \cite{taebi2019recent, ha2019chest}. The first fiducial point of interest was AO. Using the R peak indices found with Pan-Tompkins and locating the indices of the S wave endpoint, a window was defined on the SCG signal for the systole phase of every cardiac cycle throughout the dataset (red window in Fig.~\ref{fig:searchWindows}.A). The second peak within each window was located and estimated to be AO. Once AO locations were found, the temporal indices of each point were determined. The next targeted fiducial point was AC, which indicates the end of the systolic phase \cite{fukuta2008cardiac}. The peaks of the ECG T waves were located by filtering out peaks that did not meet a certain width criterion. AC and its indices were then estimated by finding the first peak within a second search window (blue window in Fig.~\ref{fig:searchWindows}.A) that was created from the ECG T wave peak to the index prior to the start of the P wave of the next ECG cycle. This data analysis process was performed for the SCG signals recorded from all three locations. Figure~\ref{fig:searchWindows}.B shows all the target CTIs and fiducial points in this study. Using the temporal indices of each fiducial point, the CTIs of interest were calculated using the equations in Table~\ref{tab:ctiFormula}.


\begin{table*}
\caption[Table]{Definition of the cardiac time intervals (CTI\NoCaseChange{s}) in this study.}\label{tab:ctiFormula}
\centering{%
\begin{tabular}{ p{1cm} p{2cm} p{12cm}}
\toprule
CTI & Equation & Description \\
\midrule
RRI & 
$t_R^{i+1}-t_R^i$ & 
RRI is the time interval between the R peaks of two consecutive QRS waveforms, and $t_R^i$ is the time instant of the $i\textsuperscript{th}$ R peak. \\

PEP & 
$t_{AO}^i-t_R^i$ &
PEP is the pre-ejection period, and $t_{AO}^i$ is the time instant of the $i\textsuperscript{th}$ aortic valve opening point on the SCG signal. \\

LVET &
$t_{AC}^i-t_{AO}^i$ &
LVET is the left ventricular ejection time, and $t_{AC}^i$ is the time instant of the $i\textsuperscript{th}$ aortic valve closure point on the SCG signal. \\

QS2 &
$t_{AC}^i-t_{R}^i$ &
QS2 is the electromechanical systole. \\
\bottomrule
\end{tabular}
}
\end{table*}


\subsubsection*{Effect of Accelerometer Location on CTIs}
To assess the accuracy of HR measurements obtained from SCG, the correlation coefficient (R\textsuperscript{2}) was calculated between the instantaneous and mean HR calculated from the ECG (gold-standard in HR monitoring) and the HR estimated from the PCG signal and SCG signals measured using all three accelerometers. Additional to the correlation coefficient between HR estimations, a regression analysis was used to evaluate the consistency between HR values estimated from different SCG measurement locations. Finally, to determine the degree of agreement between the CTIs estimated from different accelerometer locations, R\textsuperscript{2} was also calculated and analyzed between these estimations of the CTIs.

\section{Results and Discussion}
\begin{figure*}
\centering\includegraphics[width=1\linewidth]{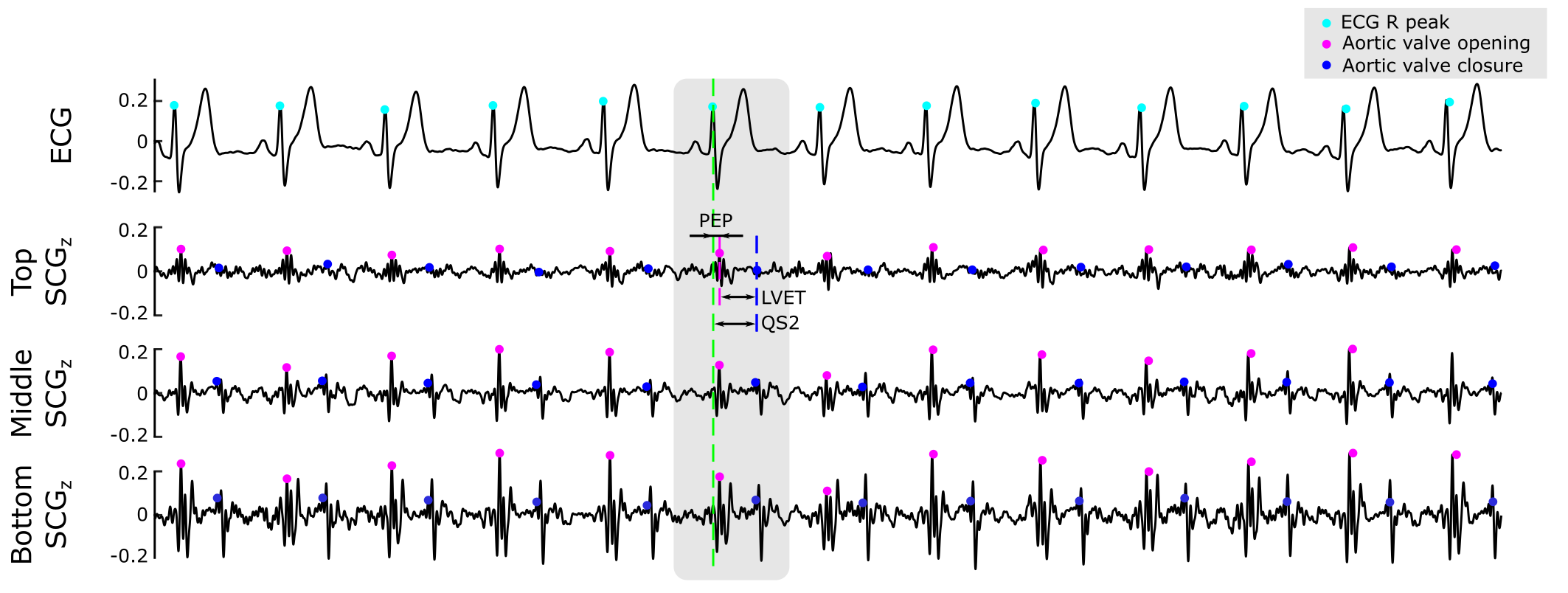}
\caption{ECG R peak and aortic valve opening and closure estimations on the SCG signals for subject 9. PEP, LVET, and QS2 estimations are shown for one cardiac cycle.}\label{fig:fiducialSub9}
\end{figure*}

\begin{figure}
\centering\includegraphics[width=1\linewidth]{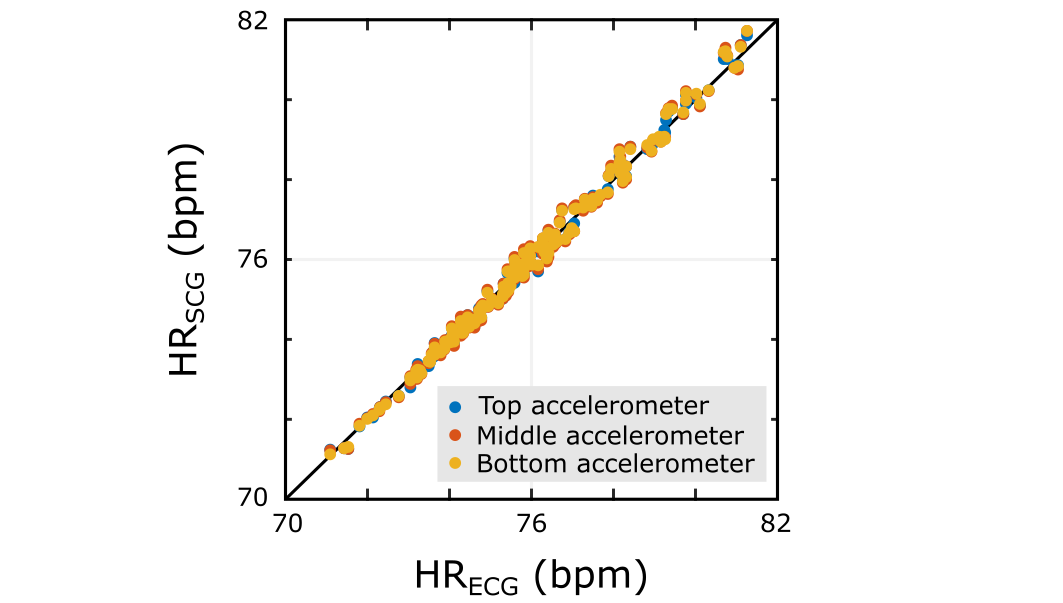}
\caption{HR\textsubscript{SCG} estimations with respect to the gold-standard HR\textsubscript{ECG} for subject 9.}\label{fig:hr}
\end{figure}

Figure~\ref{fig:fiducialSub9} shows the placement of estimated R peaks on ECG signal and AO peaks on SCG signals on a representative subject (subject 9), which were used to calculate PEP. Using the AO peaks on the SCG signals, Fig.~\ref{fig:hr} represents the HR\textsubscript{SCG} estimations from the top, middle, and bottom sensor locations with respect to the gold standard HR\textsubscript{ECG}. The correlation coefficient between HR\textsubscript{SCG} and HR\textsubscript{ECG} was larger than 0.9798, 0.9883, and 0.8329 for the top, middle, and bottom sensor, respectively, indicating the high accuracy of our SCG-based HR estimation algorithm. The complete R\textsuperscript{2} values for HR estimation on each subject is listed in Table~\ref{tab:hr}.


\begin{table}[t]
\caption[Table]{Correlation coefficient (R\textsuperscript{2}) between HR\textsubscript{ECG} and HR\textsubscript{SCG} estimated from different sensor locations.}\label{tab:hr}
\centering{%
\begin{tabular}{llll}
\toprule
Subject & Top & Middle & Bottom \\
\midrule
1 & 0.9990 & 0.9999 & 0.9997\\
2 & 0.9994 & 0.9997 & 0.9906\\
3 & 0.9853 & 0.9974 & 0.9994\\
4 & 0.9877 & 0.9883 & 0.9996\\
5 & 0.9798 & 0.9993 & 0.9666\\
7 & 0.9845 & 0.9993 & 0.9996\\
8 & 0.9973 & 0.9985 & 0.9996\\
9 & 0.9969 & 0.9953 & 0.9968\\
10 & 0.9994 & 0.9981 & 0.8329\\
12 & 0.9991 & 0.9926 & 0.9992\\
13 & 0.9857 & 0.9965 & 0.9593\\
14 & 0.9955 & 0.9936 & 0.9840\\
15 & 1.0000 & 1.0000 & 0.9999\\
\bottomrule
\end{tabular}
}
\end{table}

\begin{figure}[h!]
\centering\includegraphics[width=1\linewidth]{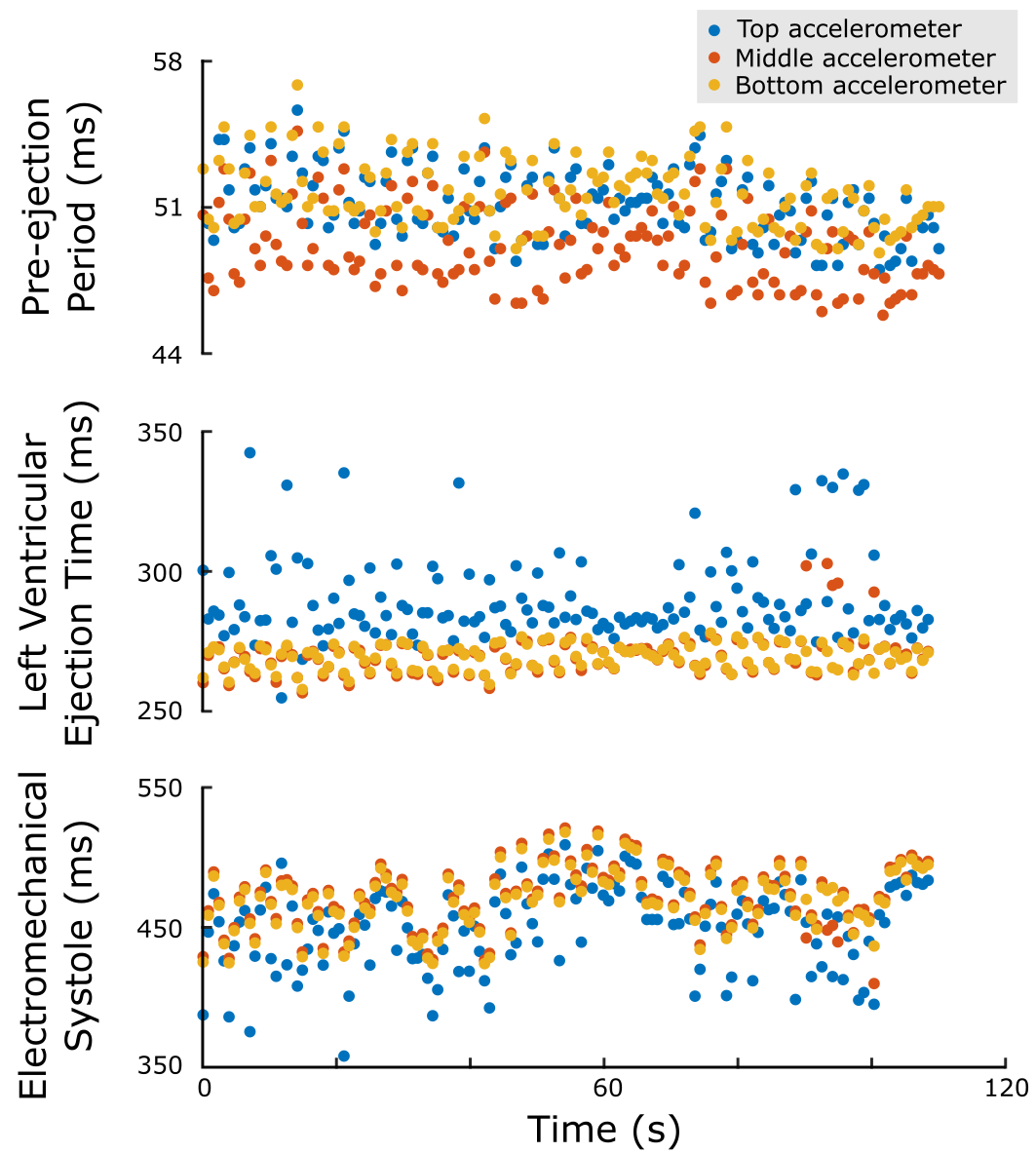}
\caption{Instantaneous pre-ejection period, left ventricular ejection time, and electromechanical systole estimated using the top, middle, and bottom accelerometers for subject 9.}\label{fig:ctiSatter}
\end{figure}

To calculate LVET, the estimated AO and AC points on subject 9 can be seen in Fig.~\ref{fig:fiducialSub9}. This figure also shows the estimated R peak to AC points for QS2. Instantaneous estimated PEP, LVET, and QS2 on each location for the same subject can be seen in Fig.~\ref{fig:ctiSatter}. Visual inspection was conducted on all signals for every subject to ensure the correct peaks  
were found.


\begin{figure}[h!]
\centering\includegraphics[width=1\linewidth]{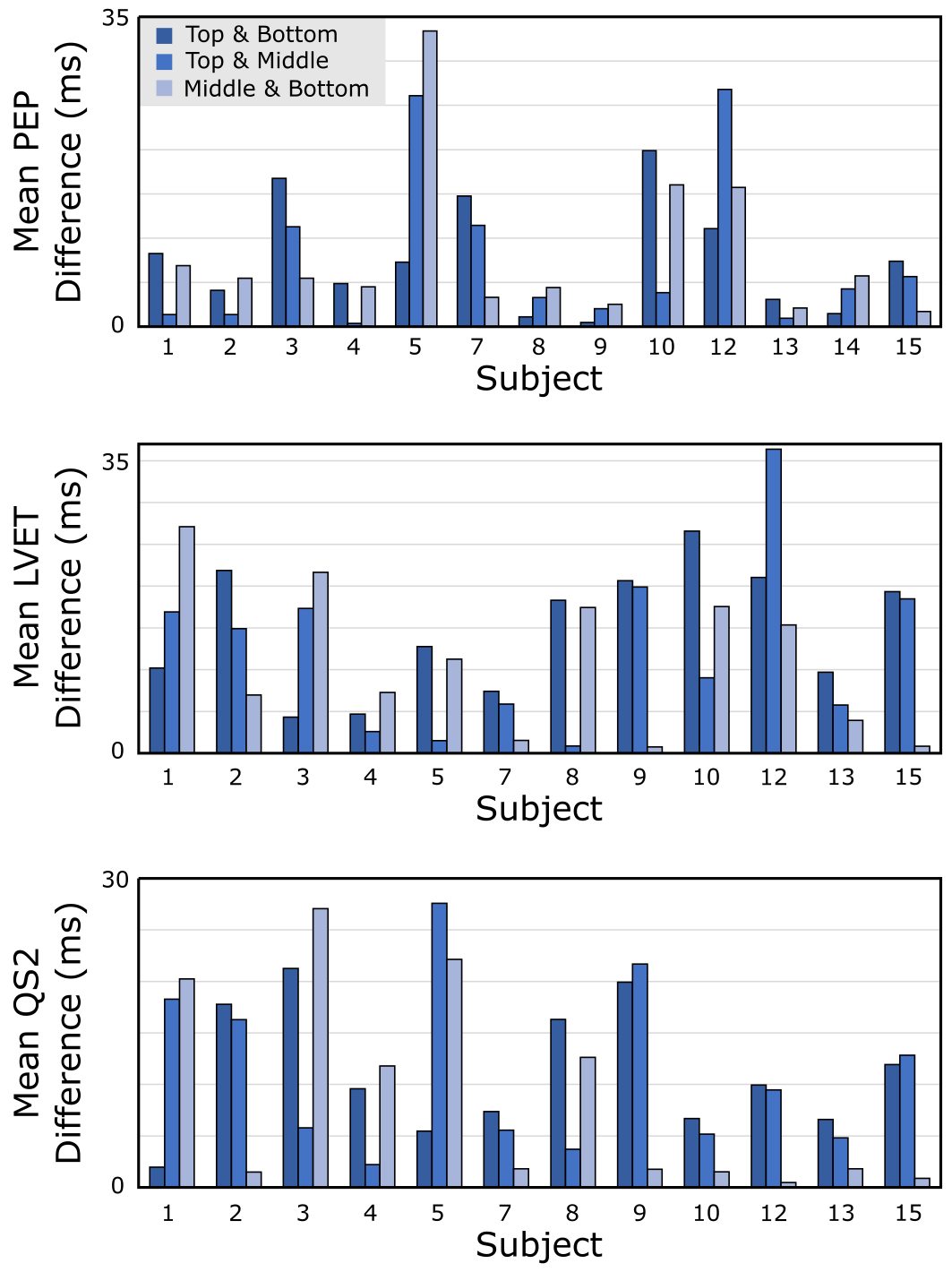}
\caption{Mean differences of the CTI\NoCaseChange{s} estimated from two sensor locations.}\label{fig:meanDiff}
\end{figure}

Due to excess noise, subject 14 was excluded from LVET and QS2. By looking at the inter-subject similarity and accuracy of heart rate for all three SCG signals, it can be concluded that aortic valve opening peaks were located precisely. By looking at Figs.~\ref{fig:ctiSatter} and \ref{fig:meanDiff}, it is self-evident that instantaneous PEP, LVET, and QS2 estimations are not the exact same for each sensor location. Yet, between the three locations, the correlation is the highest between the middle and bottom sensor. With that said, SCG signals, PEP, LVET, and QS2 estimation are dependent on chest surface location.

\section{Conclusion}

In conclusion, this study investigated the impact of sensor location on cardiac time interval (CTI) estimations obtained from SCG signals. By analyzing SCG signals collected from three different sensor locations, the study found that CTI estimations varied significantly based on the location of the SCG sensor. Additionally, the study developed a standardized approach for identifying specific windows on the SCG waveforms to obtain accurate CTI measurements across all sensor locations. The findings of this study highlight the importance of carefully selecting the location of SCG sensors when monitoring cardiovascular function, and provide a valuable methodological framework for improving the accuracy and consistency of CTI measurements. In future work, these findings will be further explored and validated using gold-standard imaging methods, paving the way for more accurate and reliable cardiovascular diagnoses.


\nocite{*}

\bibliographystyle{asmeconf}  
\bibliography{asmeconf-sample}

\end{document}